# Retina Organoids: Window into the Biophysics of Neuronal Systems

Katja A Salbaum*[1,2,3], Elijah R Shelton*[1], Friedhelm Serwane[1,2,3]

*These authors contributed equally to this work.

Affiliations
[1] Faculty of Physics and Center for NanoScience, Ludwig-Maximilians-Universität München, Munich, Germany
[2] Munich Cluster for Systems Neurology (SyNergy), Germany
[3] Graduate School of Systemic Neuroscience (GSN), Munich, Germany

Orcid IDs
Katja Salbaum: 0000-0001-5847-336X
Elijah Shelton: 0000-0001-9311-1567
Friedhelm Serwane: 0000-0001-6943-8244

## Table of contents

## Abstract

With a kind of magnetism, the human retina draws the eye of neuroscientist and physicist alike. It is attractive as a self-organizing system, which forms as a part of the central nervous system via biochemical and mechanical cues. The retina is also intriguing as an electro-optical device, converting photons into voltages to perform on-the-fly filtering before the signals are sent to our brain.

Here, we consider how the advent of stem cell derived *in vitro* analogs of the retina, termed Retina Organoids, opens up an exploration of the interplay between optics, electrics and mechanics in a complex neuronal network, all in a Petri dish. This review presents state-of-the-art retina organoid protocols by emphasizing links to the biochemical and mechanical signals of *in vivo* retinogenesis. Electrophysiological recording of active signal processing becomes possible as retina organoids generate light sensitive and synaptically connected photoreceptors. Experimental biophysical tools provide data to steer the development of mathematical models operating at different levels of coarse-graining. In concert, they provide a means to study how mechanical factors guide retina self-assembly. In turn, this understanding informs the engineering of mechanical signals required to tailor the growth of neuronal network morphology. Tackling the complex developmental and computational processes in the retina requires an interdisciplinary endeavor combining experiment and theory, physics and biology. The reward is enticing: in the next few years, retina organoids could offer a glimpse inside the machinery of simultaneous cellular self-assembly and signal processing, all in an *in vitro* setting.

## 1. Introduction

The human eye has fascinated physicists for centuries: Starting from a clump of cells, this many-body system transforms into an electro-optical device which converts photons into electrical signals to perform image processing and filtering on the fly before the signal is transmitted to the brain. Being wetware rather than hardware, it operates so differently from traditional computers in terms of hardware architecture and synchronization, and in particular much more efficiently. While the intriguing features of the *mature* retina have attracted many scientists and generated profound insight, the understanding of the physics at play in organizing early retina cells into neuronal tissues is an open field. A main obstacle is the inaccessibility of the human retina to cell tracking, biophysical characterization and electrical quantification of its network activity. Consequently, our main knowledge mainly stems from animal models.[1] As genetic edits in animal models take months to years, dictated by the growth time of the organism, linking genetic factors to the mechanics orchestrating cell organization has remained a major challenge.

This situation has changed dramatically with the advent of organoids.[2-5] Thanks to recent advances in stem cell biology, 3D *in vitro* models of the retina, called retina organoids (ROs), may now be generated in a Petri dish. Pioneered by stem cell biologists[6, 7], state of the art ROs recapitulate major cell types and tissue organization of the retina *in vivo*.[8-10] In addition, ROs show light sensitivity and early synaptic connectivity.[8, 11-13] The growth time replicates the development period of the organ *in vivo* (time scale: 20 days mouse, 200 days human), and animal testing can be avoided. Due to their *in vitro* nature, ROs grant intrinsic experimental access for recording of cell movement, genetic editing and neuronal readout. Genetic editing, in particular, is simpler compared to animal models as modified cell lines are substantially faster to generate compared to whole animals. Thus, retina organoids provide a means to study aspects of organogenesis and function normally reserved to animal models using the state-of-the-art genetic tools developed for traditional 2D cell culture.

In order to recapitulate organogenesis in organoid cultures, one must provide the necessary developmental cues. While the identification of biochemical signals has been in the hand of stem cell biologists, the characterization of biomechanical cues, including cellular forces which guide RO cell movements, is now being tackled with physical methods. Mechanical signals have been shown to impact central nervous system (CNS) development, of which the retina is part, by affecting synapse formation, axon growth and neuron positioning.[14-16] Examples of how mechanical signals influence neuronal systems *in vivo* are striking. Applying tension to axons has been shown to regulate the neurotransmitter density and thus impacts the neuronal plasticity in *Drosophila.*[17] In *Xenopus*, mesodermal stiffening initiates neural crest cell migration [18], and gradients in brain stiffness steer axon growth.[19, 20]

In this review, we will discuss RO growth and present experimental techniques capable of quantifying the mechanics of ROs. To predict how these signals lead to the formation of functional networks, theoretical models are being developed which take experimentally measured parameters as inputs. To specify critical biophysical signals through engineered microenvironments in a next step, engineers can draw upon experimental and mathematical studies. Finally, to quantify the functionality of the retina network at single neuron resolution, researchers are providing new approaches which push the limits of techniques inside the neuroscience field.[21] The reward of this inter-disciplinary approach is enticing: within the next few years, ROs could offer a glimpse inside the machinery of simultaneous cellular self-assembly and signal processing, all in an *in vitro* setting.

## 2. Retina Organoids as *in vitro* models

### 2.1 Organoids as accessible model systems

The question of how stem cells self-organize and transform into a functional retina is one of the open questions in developmental biology. In the last decade, researchers have engineered *in vitro* systems which contain the same cell types and tissue morphologies as the corresponding *in vivo* systems.[3, 5, 22] In a transformative way, those provide us with the opportunity to study (retina) development at the cell-, tissue- and organ-level *in vitro* under physiological conditions.[4, 23, 24]
Organoids are generated from embryonic stem cells (ESCs), isolated organ progenitors, or induced pluripotent stem cells (iPSCs) that self-organize into specific tissue types.[2-5] One central strategy to generate patterns of cell fate in 3D is the spatiotemporal control of transcription factors.[25] Combined with nutrients and extracellular matrices (ECMs) as well as mechanical signals, this directs organoids to recapitulate the natural development of each tissue type. The exact interplay between all biochemical and mechanical inputs during these processes is an active field of research.[24, 26]

### 2.2 Recapitulating retina development *in vitro*

The 'holy grail' of retina engineering is the growth of stratified human retina layers with full electrical functionality. Those systems could be explored for fundamental studies as well as applications such as implants, disease modelling and therapeutic testing. An exemplary milestone would be the realization of an *in vitro* retina according to specifications by the Retina Organoid Challenge.[27] Specifically, a future system should contain all major retina cell types including rods and cones, horizontal cells, amacrine cells, bipolar cells, retinal ganglion cells, as well as non-neuronal cell types, i.e. astrocytes, Müller glia, RPE cells, pericytes, endothelial cells and microglia. Ultimately, a comparison of cell types and proportions to the human eye as well as the characterization of electrophysiological activity including neuronal circuits has to be performed.

While this ambitious goal is still some years ahead, state of the art protocols generate retina which mimic the *in vivo* counterpart in some key aspects.[10, 28-30] The following sections discuss the three main steps of *in vivo* development[31] and their recapitulation using ROs *in vitro* (Figure 1). The main route to understanding development has been the identification of transcription factors (TFs) via the close observation of *in vivo* retinogenesis.[23, 24] Due to the relevance of TFs in RO protocols, the most important ones will be introduced.

## Step 1: Patterning of the eye field

The embryonic brain emerges following the definition of the main body axis (a-p). The earliest hallmark of the eye, the eye field, develops from a distinct part of the brain as depicted in Figure 1a). This tissue grows in a lateral direction in response to transverse gradients of TFs. Among the essential TFs are the homeobox-containing genes Pax6, the earliest TF conserved even across animal phyla boarders from vertebra to insects[32] and Rx[33]. Without these TFs, the eyes cannot form. Due to the role of Rx as the earliest TF specifying the eye field,[34] seminal RO protocols first have focused on the induction of Rx-positive tissues,[6, 7] rendering Rx as one of the first specific indicators of the presence of retina progenitor cells (RPCs) in ROs.[7, 35-40] With the formation of the Rx-positive eye field, a single layer neuro-epithelium, the optic vesicle, evaginates (Figure 1a). Several other TFs (including Lhx2, Six3, Six6, TII and Tbx3) are expressed,[41] which have downstream effects that pattern the neuro-epithelium into specialized cell layers.[42]

Although researchers have successfully imprinted the anterior-posterior polarity (a-p axis) on embryoid bodies,[43] ROs typically do not display such a preferred axis. Instead, distinct patches of Rx-positive cells are formed in arbitrary directions. A very recently published protocol advances the field by generating optic vesicle-containing cerebral organoids (OVB-organoids) of which 66% develop two bilateral symmetric optic vesicles (Figure 1c) from two defined primordial eye fields,[9] suggesting the establishment of a body axis perpendicular to the line connecting the two optic vesicles. To what extend this axis shares characteristics with the anterior-posterior axis of the developing embryonic brain is an open question.

## Step 2: Formation of single layer neural retina

After formation of the optic vesicle in the embryo, a second layer in differentiation hierarchy is introduced by patterning of the optic vesicle into the neural retina (NR) and the retina pigment epithelium (RPE). The NR will later constitute the retinal neuronal network. The RPE is an epithelial tissue with a protective barrier function. To set the fate of NR and RPE, a molecular switch is implemented via a combination of TFs controlling two main developmental pathways: WNT[44] and SHH[45] (sonic hedgehog), both being co-regulated by a third TF (VSX2 visual system homeobox 2, also referred to as CHX10).

***Specification of cell fate***. The activation state of WNT triggers retinal progenitor differentiation into either NR (WNT signaling inhibited by VSX2 [10, 46, 47]) or into RPE (WNT signaling active [6, 7, 47]). As the decisive factor between NR and RPE is VSX2, it is widely accepted as an indicator of cells with NR identity in ROs as well.[9, 35, 37-40, 46] RPE fate is co-controlled by another TF (MITF) that it is expressed throughout the entire structure in the early optic vesicle. Later on, it is restricted to the RPE.[44, 48, 49]

After optic vesicle growth, a transformative morphological step follows: its invagination to form the optic cup which results in the curved morphology of the retina *in vivo*. Recapitulated in organoids, [6, 7, 9, 50] the efficiency varies greatly, depending on the protocol and epigenetic memory of the cell line [46, 50, 51]. While the formation of a true optic cup is extremely rare,[46, 52] the laminar organization of the NR forms reliably in ROs, even without the formation of an optic cup,[8, 35, 38, 39] if a cell line that is able to build ROs is used.

While the role of biochemical factors has been predominantly studied, the role of external mechanical signals for optic vesicle development is poorly understood. Recent *in vivo* experiments indicate the requirement of a mechanical coupling between the optic vesicle and the adjacent tissue.[53] This mechanical link is established via the extracellular matrix (ECM) and is crucial for the proper formation of neuronal connections.[42, 53-55] In RO protocols, a direct mechanical impact is implemented, e.g. via the excision of the optic vesicle,[6, 7, 40] or the attachment of the embryoid body (EBs) onto planar substrates. To this end, 3D EBs as described in Figure 1b, are plated onto an ECM-coated surface and grow in '2.5D'. When optic (VSX2-positive) vesicle-like structures appear, they are lifted manually and left to mature as 3D organoids in floating culture.[8, 37-39, 56] At the same time, ROs can also be generated without any direct mechanical interaction,[36, 57-61] emphasizing that the impact of mechanical signals is still an open question.

In the future, the controlled application of mechanical signals might allow for optimized organoid growth.[62, 63]

### Step 3: Development of a functional layered organization

Once the decision between NR and RPE is made, the NR tissue undergoes drastic changes in a final step by transforming from a single layer epithelium into a multilayer tissue.[64] The physics of this assembly is rich: Retina progenitor cells begin to stretch across an NR cell layer.[64, 65] Their cell body begins to oscillate between apical and basal side with an oscillation frequency that is coupled to the mitotic cycle. Once the cell body arrives at the basal side, the cells are more likely to switch from a proliferative to a postmitotic stage and start to differentiate. In this way, the retina progenitor cells (RPCs) differentiate further into the seven retinal cell types[34] fulfilling different functions:[66] rod and cone photoreceptors, bipolar cells, retinal ganglion cells, horizontal cells, amacrine cells and Müller glia. The differentiation process is then followed by the migration to their final position. Finally, synaptic connections to their intended neighboring neurons are formed.[64]

Several cell type specific TFs that determine cell fate specification have been investigated on a genetic and transcriptomic level,[8, 67-72] and reviewed in depth.[73-77]

In RO culture, the layers of the NR can be generated, although the curvature of the layers is opposite to the *in vivo* version. Some cell lines are also unable to form ROs[46, 51, 78-80] which is suspected to be due to epigenetic preprogramming dependence on the stem cell source.[50] On the other hand, if the cell line is able to form ROs, it displays the complete set of major cell types found in the retina *in vivo* and also has comparable expression profiles.[8]

For the photoreceptors to be functional, the culture medium must be enriched with factors that would be provided by RPE *in vivo*. Therefore, retinoic acid and taurine are part of almost all published protocols.[10] In some cases, WNT/BMP inhibitors are added leading to increased retinal progenitor differentiation, confirming the biochemical patterning mechanisms described *in vivo.*[46]

An RPE layer directly in contact with photoreceptors acts as a protective cushion *in vivo*. This interface has not been reproduced by recent human iPSC RO protocols.[8, 46] Instead RPE grows as single layer epithelial patches spatially separated from the NR (Figure 1c). To overcome this problem, bioengineered approaches have been realized to form RPE-NR contact.[81, 82] Another form of organoids that have an RPE layer attached to photoreceptors are optical vesicle-containing brain organoids (OVBs) (Figure 1c). The current drawbacks of OVBs is their less organized laminar structure compared to established protocols of ROs and the NR *in vivo*.[9] Therefore, the challenge of generating highly self-organized ROs with a physiological layer of RPE is not yet solved.[27]

While post-natal light deprivation of mammals hinders the development of the brain areas that processes visual inputs,[83-85] it has not been shown that the pre-natal functionalization of the retinal layers is light-dependent. Whether RO show similar optical properties, including the intriguing light guiding features provided by photoreceptor nuclei and Müller cells as observed in the retina *in vivo* has yet to be shown[86-88].

## 2.3 Neuronal readout of retina organoid function

Photoreceptors transform light into an electric signal which is then conveyed mainly via bipolar cells to the retinal ganglion cells (RGCs). The RGCs relay this signal to the brain via the long axons that form the optic nerve. Beyond this simple picture, the retina features a rich variety in retinal signal processing.[89, 90] Neuronal readout performed in animal models[91] has triggered computational models for retina neuronal network activity.[92, 93] Today, more and more retina circuits are being revealed[94] and are an open topic in neuroscience.

Organoid technology and experimental neuroscience tools are advancing hand in hand to overcome one major challenge: recording neuron activity at millisecond time resolution in 3D human neural tissue. Existing techniques differ in their readout speed, their applicability (2D vs 3D) and their resolution (subcellular, cellular, tissue level). An excellent comparison of technologies is provided by Passaro et al.[95]

with a focus on cerebral organoids. In particular, recent work has shown functional networks in cerebral organoids via a combination of recording technologies.[21, 96]

In ROs, however, the existence of major neuronal circuits has not been reported so far[97] and is currently being tackled via different approaches. While traditional patch clamp experiments have offered unprecedented timing information on individual RO neurons,[39] it prevents the readout of a large set of neurons organized in a network. Parallel stimulation and readout of thousands of retinal neurons using multielectrode arrays has reported light sensitivity of RO cells.[11-13] However, the 2D architecture of the electrode arrays prevents measurements of the 3D retinal tissue. In addition, the missing link between the electrical recording and the neuron type poses a hurdle on the characterization of circuits. The combination of single cell patch clamping and averaged readout of network activity via electroretinography (ERG) has shown activity changes upon optical stimulation.[9] The parallel imaging of neuron activity in a network has been realized using 2-photon fluorescence microscopy of calcium indicators. Recordings of activity upon optical stimulation of cells from the outer and the inner nuclear layer have revealed the existence of synaptic connections[8] in a small fraction of neurons (~ 10%). However, these measurements have been performed within a single plane (2D) due to the limited acquisition speed of the confocal microscope. Thus, the mapping of 3D retina networks still remains a challenge. While recent experiments have mapped neuronal networks in cerebral organoids,[21] such studies in retinal organoids are still lacking.

## 3. Physics of Retina Organoids

How does a many-body system composed of neurons self-assembles so that it can perform computational tasks? On a microscopic level, the formation of neuronal networks relies on cellular behaviors such as proliferation, differentiation and migration, to first organize cells into tissues which later develop into neuronal networks. As cellular behavior is controlled via biochemical signals which cells receive from their environment, their identification has been the building block of developmental biology. The innovation of biophysical tools, however, has led to the discovery that cells also sense physical cues to reconfigure their behaviors. Even the differentiation of cells is regulated by the stiffness of the underlying substrate as seminal experiments using *in vitro* cell culture have shown.[98] As the field of mechanobiology matures,[26] the brain now becomes a target of inquiry. How is the formation and function of neuronal networks, as, for example, in the brain, controlled by mechanical signals? Today, it is known that mechanical signals impact key neuronal behavior including axon formation, migration, plasticity and synapse formation.[15, 16, 99, 100] ***On an organ scale***, brain architecture is hypothesized to form via an interplay between chemical and mechanical signals as reviewed in Llinares-Benadero et al.[101] The central idea is that brain folds emerge due to differences in mechanical properties and/or forces within distinct cerebral layers.[102, 103] The mechanical cues, in turn, are regulated by and influence biochemical signals which, for example, control cell proliferation leading to pressure changes or alter extracellular matrix composition causing changes in elastic modulus. ***On a single neuron level***, researchers demonstrated that neuron branching is controlled via the stiffness of the underlying substrate.[104] One of the most important mechanism of the mechano-electrical coupling are force-sensitive ion channels such as Piezo1 and Piezo2. Their activation by mechanical force triggers $Ca^{2+}$-signaling cascades to influence a variety of cell behaviors, including differentiation.[19, 99, 105, 106] Accordingly, their discovery[107] has been awarded with the Nobel Prize in Physiology or Medicine 2021.

***From single neurons to networks***

*In vivo* biophysical studies of network formations are rare due to the limited accessibility of the organism's nervous system to mechanical manipulation, cell tracking and simultaneous neuron readout. The few studies which have been performed have strengthened the notion that mechanical forces guide neuronal network development also *in vivo*. When applying tension to neurons in developing *Drosophila*, researchers induced the clustering of neurotransmitters at presynaptic terminals.[17] Tuning the synaptic strength by forces could have a tremendous impact on neuron function. Similarly, stretching axons in living *Drosophila* resulted in the accumulation of synaptic vesicles.[108] The tissue's elastic modulus determines the rate and direction of axon growth of retinal ganglion cells and also controls the axonal branching in the optic tectum of developing *Xenopus*.[19] Dynamic recordings of changes in the gradient of tissue elastic modulus revealed its role in controlling key events in axon pathfinding[20]. Even the migration of neural crest cells has been found to be controlled by the stiffness of the underlying mesodermal tissue.[18]

Growing evidence from *in vivo* animal models and *in vitro* single cell studies suggests that mechanical signals play a key role in neuronal system development.[15, 16, 99, 100] However, the direct observation of how mechanical signals impact neuronal function within the CNS is still lacking.

***Retina organoids as a window to the brain***

For centuries, the eyes have been called the *window to the soul*; for researchers today, the retina offers a window to the brain. As part of the CNS, the retina shares many properties including functionality, response to injury, immunology, and basic cellular (patho-)mechanisms with the brain[109]. Therefore, studies of RO development and neuronal network function might provide insights into the general role of mechanics in CNS development and function. This opens the door for experiments and theory to develop a physical model of neuronal tissues which integrates both mechanical and electrical aspects. It could solidify the link between the fields of mechanics and electrics in biological systems.[110, 111]

A biophysical understanding of RO development requires a close collaboration between experimental and theoretical approaches (Figure 2). Experimental techniques to probe mechanical properties and quantify forces need to be implemented (Figure 2a). The measurements then provide inputs for mathematical models of 3D multicellular systems (Figure 2b). Those, in turn, make testable predictions for organoid shape development and growth and informs bioengineering approaches (Figure 2c).

This section provides an overview of mechanobiology tools which have been applied to retina as well as retinal and cerebral organoids. Different mathematical models for the morphogenesis of CNS (neural and retinal) organoids are presented.

## 3.1 Experimental mechanobiology tools

The mechanical signals that neurons experience in a network depend on two components: first, the internal and external forces acting on and within the neuronal network and, secondly, the tissue mechanical properties which dictate how those forces are transmitted across several neurons. Direct measurement of mechanical properties requires controlled application of stress while recording the material deformation, i.e. the material strain, or vice versa. In the last decades a range of biophysical tools has been developed to quantify cellular forces and the mechanical properties at various time scales and length scales.[112-115] The length and timescale of the application of stress (strain) play a crucial role: quantifying the mechanical properties of the local environment at cellular and tissue levels requires probing the system at a length scale typically on the order of 10 microns and larger. Furthermore, many relevant biological phenomena, including synapse formation within neuronal networks, occur at time scales of minutes to hours.[116] Thus, performing biomechanical measurements relevant for these phenomena requires techniques that probe the system at those scales.

### *Atomic force microscopy*

Atomic force microscopy (AFM) measurements require the tip of a cantilever of known mechanical properties to be brought into contact with the sample surface.[117] In this way both force and displacement may be monitored. Indentation experiments using AFM have found spatially varying elastic moduli in different regions of the guinea pig retina.[118] Applied to mouse ROs a difference in elastic modulus of the two initial tissue types, RPE and NR has been revealed with the RPE tissue being two-fold as stiff as NR (400 Pa vs 200 Pa).[6] This difference in tissue stiffness has served as input for vertex models which have provided insights into optic cup morphogenesis.[6, 119] The mechanism of cup formation is proposed to work in the following way: Upon differentiation of the optic vesicle into NR and RPE, the reduction of acto-myosin concentration causes the bending of the NR. This, in turn, leads to the increase of tissue strain at the intersection between NR and RPE. Consequently, strain-triggered Ca-release activates lateral constriction of the NR which initiates optic cup formation.[119] *In vivo* AFM revealed a stiffness gradient in the embryonic *Xenopus* brain which controls the growth direction of retinal ganglion cell axons.[19] State of the art experiments allow the simultaneous mapping of mechanical properties and imaging the dynamics of axon growth *in vivo*.[20]

Since AFM relies on surface interaction with the sample, this method remains limited to 2D surface measurements and does not provide measurements inside intact 3D tissues (*in situ*). In addition, the simultaneous neuronal readout required for the quantification of the electrical network remains a challenge due the spatial constrains the AFM imposes on the experimental setup.[117]

### *Brillouin microscopy*

Brillouin microscopy, reviewed in Prevedel et al. 2019,[120] harnesses the excitation of acoustic phonons via photons within the tissue. Here, the timescale of measurement is dictated by the GHz frequency shift of the Brillouin scattering event. Confocal Brillouin microscopy has been introduced by Scarcelli at al. 2017,[121] and has then been applied in the context of nervous system development while imaging the mechanics of spinal cord development in zebrafish.[122]

Brillouin microscopy and AFM based measurements on retina explants showed an agreement between the measured frequency shift of Brillouin microscopy and the elastic modulus determined by AFM measurements.[123] In addition, a gradient in the elastic modulus across the retina layers was revealed. When applied to mouse retina explants, recording the Brillouin frequency shift makes it possible to distinguish between the different retina layers.[124] Since Brillouin interrogates on nanosecond timescale, it is challenging to extrapolate the mechanical responses from second to hour timescales which are often physiologically relevant, e.g. in the formation of synapses.[116]

### *Laser microdissection*

Precise microsurgery of tissues has been performed using pulsed lasers.[125] Ablating cell-cell junctions in a controlled way and recording vertex movements as a function of time has provided estimates for line tensions to serve as inputs for the vertex model.[126-128] When applied to developing ROs, laser ablation of cells in the intermediate region between NR and RPE has provided qualitative insights into cell proliferation rates in the regions of the NR and RPE and the structural link between those.[6] Beyond probing tissue mechanics, laser ablation has induced calcium transients and revealed that the local up-regulation of intra-cellular calcium concentration drives lateral constriction and thus contributes to the formation of the optics cup.[119]

Ablation of Müller Glia cells in the zebrafish retina revealed their role as mechanosensitive springs holding the retina together in a compressive way.[129] Recently, the role of mechanical interaction of the developing eye field with the adjacent tissue has been probed *in vivo*[53] with the finding, that a mechanical contact of the forming optics vesicle to the adjacent tissue is required for proper development. Beyond mechanical manipulation, the technology has been used to characterize neuronal circuits in the zebrafish brain[130] and has revealed that the mechanosensitive ion channel Piezo1 also controls neuron regeneration.[131]

Laser ablation is a powerful tool for introducing mechanical perturbations to cell and tissue mechanics. In order to use it as a quantitative probe for the elastic modulus, further assumptions for cell and tissue mechanics have to be made.[132]

### *Optogenetic manipulation*

Optogenetic tools allows the exposure of certain wavelengths of light to trigger the expression of specific genes. To assess how myosin-dependent active cellular forces contribute to organoid formation, researchers have developed an optogenetic method to reversibly control the active myosin concentration by triggering a light-sensitive molecular switch for apical myosin activation (*Optoshroom3*).[133] Importantly, the optical activation of cellular forces has allowed to reversibly change tissue shape. This permits to isolate the role of mechanical forces for shape formation of the organoid.

### *Ferrofluid microdroplets*

To allow in situ measurements of mechanical properties at the timescale of minutes, ferrofluid droplets[134, 135] could serve as a complementary technology to AFM and Brillouin microcopy. Ferrofluid droplets can be microinjected into different retina layers of the organoid model using a pressure injector. To control the length scale of measurements the droplet diameter can be adjusted between 30 and 100 microns by tuning the injection time and/or the injection pressure, or volume if a volume displacement hydraulic injector is used. Controlled mechanical stresses can be applied locally to the tissue by a homogeneous magnetic field which causes the droplet to elongate along the magnetic field axis. Recording the droplet's dynamical shape as a function of time and fitting a viscoelastic model allows the quantification of tissue mechanical properties, including its elastic modulus and its viscosity. While measurements at timescales from seconds to minutes are possible, the drawback of the technique in its current state is the low throughput. In addition, the spatial resolution of the mechanics measurement is supra-cellular as it is linked to the droplet diameter.

### *Implications for organoid bioengineering*

Advancements in mathematical modelling and biophysical experimental tools will work together to enable better bioengineering of organoids. Understanding the interplay between mechanics and development allows one to appropriately specify external mechanical signals, such as the stiffness of the environment, which then can be used to engineer organoid growth.[136] A prime example is the micropatterning of the cerebral organoid environment mechanical using a microfluidic organoid chip by Karzbrun et al.,[137] which recently allowed the reproducible recapitulation of neural tube folding in such a biochip.[63]

Besides mechanical cues, the chemical signaling of the neuronal network with its surrounding tissue is an import aspect to design. The majority of recent RO protocols has focused on the generation of layered retinas,[10] but with some exceptions, payed less attention to the interface of the RPE with the photoreceptor layer.[6, 7, 9] Previous studies have shown that the RPE layer is important for retina homeostasis as well as the integrity and maturation of photoreceptors.[138] To overcome this problem, Achberger et al.[81] have developed a retina chip based on microfluidics which interfaces a layer of iPSC-derived RPE cells and ROs. Such developments affirm that the future of RO biology depends on continued interdisciplinary engagement between experimentalists, theorists, physicists, biologists, and engineers.

## 3.2 Mathematical modeling

So called *in silico*, or mathematical, models are valuable tools for understanding the key principles behind the self-assembly of organoids and the organs after which they are modeled. Accordingly, several models have been developed which differ in their complexity and coarse graining level, as reviewed before.[139, 140] Models vary both in the kind of biophysical phenomena simulated, as well as the level of detail at which the system is resolved. These range from simplistic reaction-diffusion based models,[25] which only consider biochemical factors, to 3D multicellular mechanics model which describe tissue mechanics at a single-cell level and include a chemo-mechanical coupling.[141] Here, we restrict our focus to biophysical models which simulate the mechanical forces and rheology underlying retina organoid morphological organization. Within this constraint, we can consider the problem across various levels of detail, from fine-grained agent-based models (vertex models) to coarse-grained equation-based models (shell models). When evaluated in conjunction with experimental measurements of biophysical parameters, mathematical models can inform bioengineering efforts which seek to recapitulate organ formation and function in experimental organoid models more comprehensively.

### *Vertex models*

To account for the dynamics of individual cells, more fine-grained, and computationally demanding, agent-based models have been developed. Vertex models[126-128] are agent-based models in which the network of cells forming a tissue is represented by a set of points common to neighboring cells. Those are termed vertices. Virtually all vertex models are overdamped, meaning inertia is neglected and the velocities of vertices are proportional to the net forces. Consequently, the force acting on a single vertex is written as

$$F_i = \xi \frac{dx_i}{dt}$$

where $F_i$ denotes the total force acting on vertex $i$ located at position $x_i$. $\xi$ is the viscous friction coefficient and $t$ the time. Models are distinguished from one another mainly by how the forces are derived from the state of the network. For example, the force may be related to an energy functional which depends on the line tension as well as the hydrostatic pressure related to the geometry of the cell (area, volume). Okuda and coworkers[142] incorporated additional viscosity terms in the vertex model, i.e. the friction force acting on each vertex is proportional to the difference in the local vertex velocity vectors and the relative velocities of surrounding vertices. While vertex models have traditionally been applied to 2D cell monolayers they have been further developed to 3D systems.[142-144]

3D vertex models are successful in describing epithelial sheet morphologies and dynamics, such as the super-elastic mechanical behaviors observed in doming of cultured MDCK cells.[145] Building on the work of Krajnc et al.,[146, 147] researchers have shown that 3D vertex models reproduce many geometries found in the organoid field.[148] Most recently, Yang et al.[149] used 3D vertex models to understand how force patterning leads to mechanical compartmentalization in intestinal epithelium. While providing physical insights into cell sheet dynamics, these models are limited to a single layer epithelium in 3D. The particular challenges to advance to 3D modelling mainly is the additional symmetry breaking via the apico-basal polarization of cells and the contact with the extracellular matrix.[150]

One of the most advanced theoretical models for organoids[119] couples the 3D vertex model[142] with biochemical signaling based on a reaction-diffusion mechanism.[25] The chemo-mechanical coupling is implemented in the following way: the cell proliferation rate is linked to the concentration of an activator molecule. The local increase in division rate leads to pressure increases and thus triggers tissue scale morphogenic movements. The model recapitulates various behaviors on a cell scale, such as deformation, re-arrangement, division, apoptosis, differentiation, and proliferation and, on a tissue scale, undulation, tabulation and branching.

The model has been applied to ROs and has been fed with experimental parameters.[119] The simulations are consistent with a mechanism where regions of the optic vesicle independently differentiate into NR and RPE. Both regions show differences in mechanical properties (elastic modulus) and pressure resulting from cell division. Consequently, a deformation is generated at the intermediate section between NR and RPE, which in turn triggers mechanosensitive ion channels on the basal side. This ultimately leads to a lateral constriction which facilitates NR invagination.[119]

The description of the organoid as a network of polygonal (for 3D, polyhedral) shapes in combination with simple mechanical rules makes the vertex model a powerful tool to understand and predict organoid remodeling. However conventional vertex models focus on a single layer epithelium undergoing morphological changes in 3D, and it is not obvious that the cell interactions modelled map neatly onto the cell interactions within the retina or ROs. Retinal tissues are neither epithelial, nor single layer, and there is clearly room to address these shortcomings in future cell-level models.

### *From vertex models to continuum descriptions*

Vertex models derive tissue morphologies from the mechanical interactions between cells. On the one hand, this approach can help to associate particular morphologies with individual cell behaviors. One the other hand, these models require many specified parameters. When the parameter space is adequately reduced, the vertex model allows conceptual breakthroughs.[151] For this reason, an effective strategy is to find ways to reduce such agent-based simulations to minimal set of features which map onto the various morphological regimes of interest. In this way, a conceptual understanding is built how tissue morphologies arise from detailed cell behaviors as well as effective continuum, tissue-level descriptions.

Recent studies bridge the divide between discrete cell-level simulations and continuum models. Vertex models of stretching epithelial tissues were used to derive the effective superelastic continuum mechanical response experimentally observed in doming MDCK cells.[145] In another study with epithelial tissue, the authors tackle crypt formation in intestinal organoids. They first scan a parameter space defined by vertex models and then use the results to distinguish between different mechanism by deriving scaling laws associated with morphometrics for each mechanism.[149] Using finite element methods while neglecting cellular structure, researchers have modeled the optic vesicle and the adjacent surface ectoderm as a thin elastic shell with patterned mechanical properties.[55] To constrain the parameter space, the work has integrated mechanics and organoid shape measurements as inputs to the simulation. As a result, the connection between optic vesicle and underlying surface ectoderm has identified as crucial for optics cup morphogenesis, raising the question how external mechanical signals from other tissues are required for proper retina development.[42, 53-55]

There are several benefits of mapping cell-level models onto continuum descriptions. One benefit is experimental testing of models. While the parameters which feed into vertex models may not be directly observable, physical properties such as elastic moduli (stiffnesses) can be experimentally determined. Additionally, continuum descriptions can be adapted to scenarios which vertex modelling frameworks cannot be easily extended. For example, cell-cell interactions among the multiple layers of the retina may be poorly approximated within a single layer epithelium model, but mapping such a model to a continuum level description may provide meaningful insights. The approach to modelling epithelial tissues could provide insights for how to develop equation-based models for retinal organoids and neural organoids which aim to explain optic cup formation and surface wrinkling. In the next section, we consider the use of shell models, a type of continuum model, to explain wrinkling morphologies observed in the brain and neural organoids.

### *Shell models*

The particular strength of analytical shell models is that they successfully describe tissue-scale wrinkling morphologies in organoids with a minimum set of parameters. Here, the organoid is modeled as a bi-layered material consisting of a core and a surrounding shell. Differences in the mechanics, i.e. the pressure and/or the mechanical properties of the shell with respect to the core, lead to the emergence of wrinkles at characteristic length scales according to elastic bilayer instability. Those folding instabilities have been found to play an important role in brain development[101-103] and have been recapitulated in seminal experiments with cerebral organoids.[137] Upon growth, the organoids form wrinkles characterized by a length scale $\lambda$. Several shell models have been developed which differ in assumptions for the mechanical properties and/or pressure differentials of the core and the shell.[137, 152-154] In,[137] the observed organoid forms cusped folds, consistent with an analytical model in which the shell has a lower elastic moduli and higher swelling rate than the core.[102, 103] The analytical model predicts a linear scaling of wrinkling wavelength with respect to shell thickness, namely,

$$\lambda = 2\pi\, t \left[\frac{\mu_s}{3\,\mu_c}\right]^{1/3},$$

where $t$ denotes the shell thickness and $\mu_s$ and $\mu_c$ are the elastic moduli of the shell and core, respectively. This linear scaling behavior has been experimentally observed.[137] In a complementary model[152] of the same experiment, the organoid is described as an elastic core covered by a fluid-like film which is held together by springs. Such a model could find application in the retina as Mueller glial cells have identified as springs holding together the retinal tissue.[129, 152] In ROs, Müller glial cells have been found to span the entire NR.[23, 46] Applying this so-called "buckling without bending" model[111], the authors were able to reproduce oscillations in the core deformation amplitudes which are out of phase with oscillations in shell thickening.

While the physical mechanisms by which wrinkling arises substantially differs from other descriptions,[137] this model reproduces shapes observed in both cerebral organoids and retina layers.[152]
Shell models can be made more complex by considering fluid-like relaxations[154] or surface tension effects,[153] with the aim of more accurately reproducing observed morphologies. Adding additional parameters can provide insights into the mechanical states of the tissues, or reflect cell-scale features which cannot be directly modeled. While most models predict linear scaling of wavelength with respect to film thickness, more complex models are distinguished by their predictions for the onset of the instability and the scaling as a function of stiffness gradients.
The beauty of shell models is the minimal set of parameters they are based on which promises a fundamental understanding of relevant physical factors. The models generate testable predictions such as the scaling of local curvature with the thickness of the neuronal shell of the organoid. Although most models have been developed and applied in the context of brain and neuronal organoids, they could be transferred to ROs.[152] In their current state, however, they are too coarse grained to account for individual retina layers. Shell-based analytical models capture the shape variations including shell thickness and local curvature in neural organoids. However, several models with different mechanical properties and pressure differentials are compatible with the experimental observation of shape formation in neural organoids. Therefore, quantitative measurements of mechanical parameters are required which will allow to discriminate between the models to generate a unique picture of the underlying physical principles. Modelling approaches which make clear predictions of mechanical properties and/or forces become testable as experimental biophysical tools become accessible.[134, 135]

## 4 Outlook

The engineering of human neuronal networks in a Petri dish would be a prime example of Richard Feynman's saying: *What I cannot create, I do not understand.* In this sense, the recent engineering of cerebral organoids might provide basic information about brain functioning. Intriguingly, the systems have shown the emergence of synchronized neuronal network activity [21] without external sensory inputs. At the same time, cerebral organoids are highly complex regarding their genetic background, their organization and their number of cell types.[12] This raises the question whether simpler model systems exist which still mimic key functionality. The retina, as part of the central nervous system, has some well characterized circuits[89, 90] and closely mimics the tissue organization *in vivo*.[8] The building up of functional networks in ROs by controlling biochemical and mechanical parameters, could provide a complementary approach to understand the self-assembly and function of neuronal systems. The system's unique benefits are the lower complexity compared to cerebral organoids while some of the most fascinating aspects of the central nervous system are maintained.

Within the next years, a mechanical understanding of retina development could be in experimental reach. The recording and the theoretical modelling of the *electrical* signals, however, will be a formidable challenge.

## Acknowledgements

We thank Oliver Drozdowski and Ulrich Schwarz for helpful comments on the manuscript. This project has received funding from the European Research Council (ERC) under the European Union's Horizon 2020 research and innovation program (grant agreement n° 850691), the Center for NanoScience (CeNS) at the Ludwig-Maximilian-University (LMU) Munich and the Munich Cluster for Systems Neurology (SyNergy).

## Conflict of interest

The authors have no conflicts to disclose.

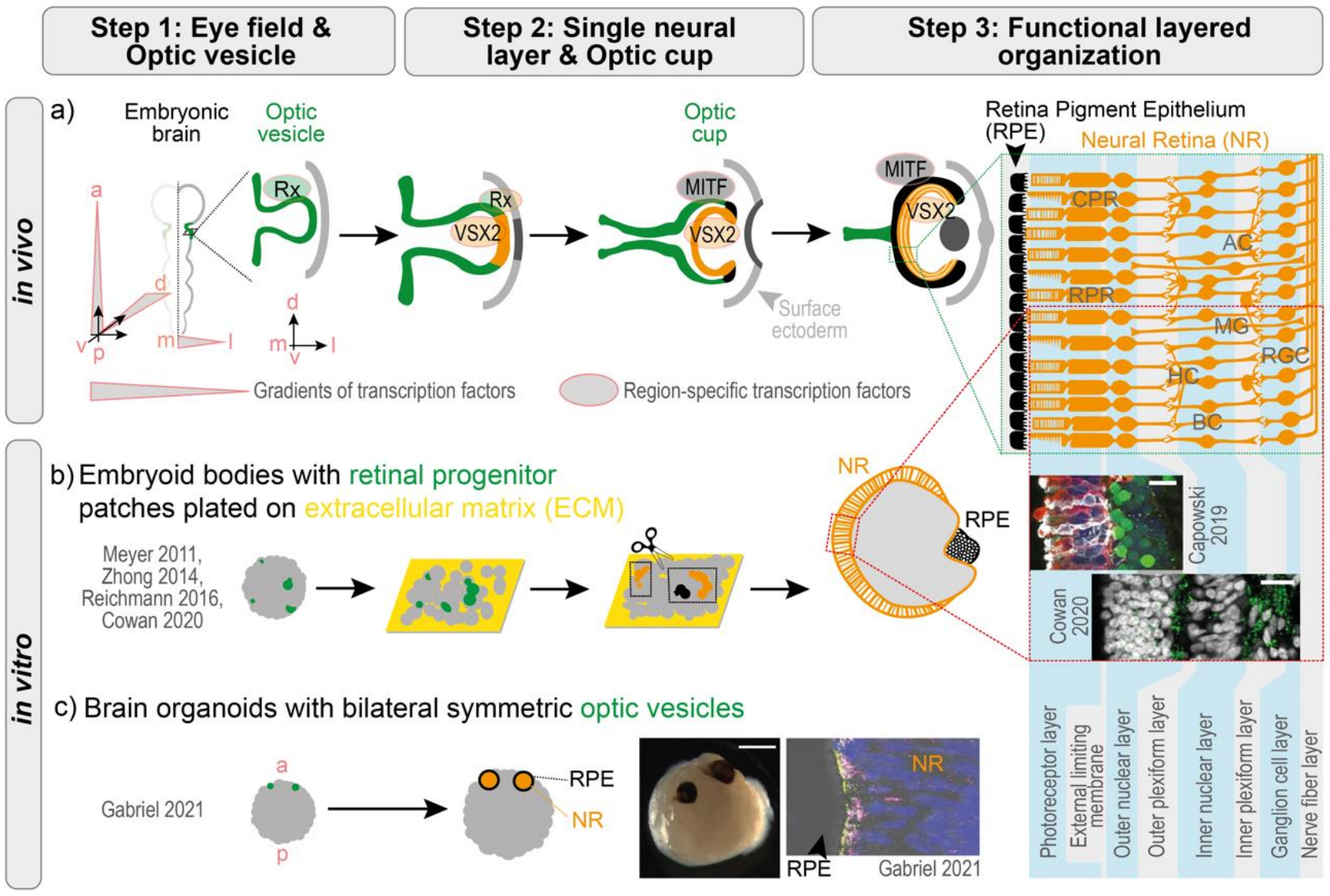

*Figure 1:* ***Retina Organoids (ROs) recapitulate key aspects of eye development. a)*** *In vivo, eye formation is initiated by gradients of biochemical signals which define the three body axes (a: anterior to p: posterior; v: ventral to d: dorsal; m: medial to l: lateral). Starting from the embryonic brain, the eye field forms laterally to develop a vesicle-like shape (optic vesicle). This undergoes a morphological transformation into the optic cup imprinting the retina shape. Specific transcription factors (Rx, VSX2, MITF) set cell fate to retinal progenitors (green), neural retina (NR, orange) and retinal pigment epithelium (RPE, black). Surface ectoderm is indicated in shades of grey. The functional neural retina (step 3) consists of the following cell types: RGC: retinal ganglion cells; RPR: rod photoreceptors, AC: amacrine cells, BC: bipolar cells; CPR: cone photoreceptors; HC: horizontal cells; MG: Müller glia.* ***b)*** *Established protocols build on spontaneously differentiating regions of retinal progenitors (green). After contacting the extracellular matrix (ECM, yellow), those mature into NR and RPE recapitulating all cell types of the NR with patches of RPE.* ***c)*** *Optical vesicle containing organoids (OVB) recapitulate anterior and posterior regions. The optic cup-like structures are formed more reproducibly compared to b) and display a more physiological connection between RPE and NR. However, the organization of the NR is less developed compared to protocols used in b). Scale bars: 20 microns (b), 1mm (c). Reproduced from Capowski et al., Development 146, 1 (2019)[46], licensed under a Creative Commons Attribution (CC BY) license; Cowan et al., Cell 182, 6 (2020)[8], licensed under a Creative Commons Attribution (CC BY) license; Gabriel et al., Cell Stem Cell 28, 10 (2021)[9] with permission from Elsevier.*

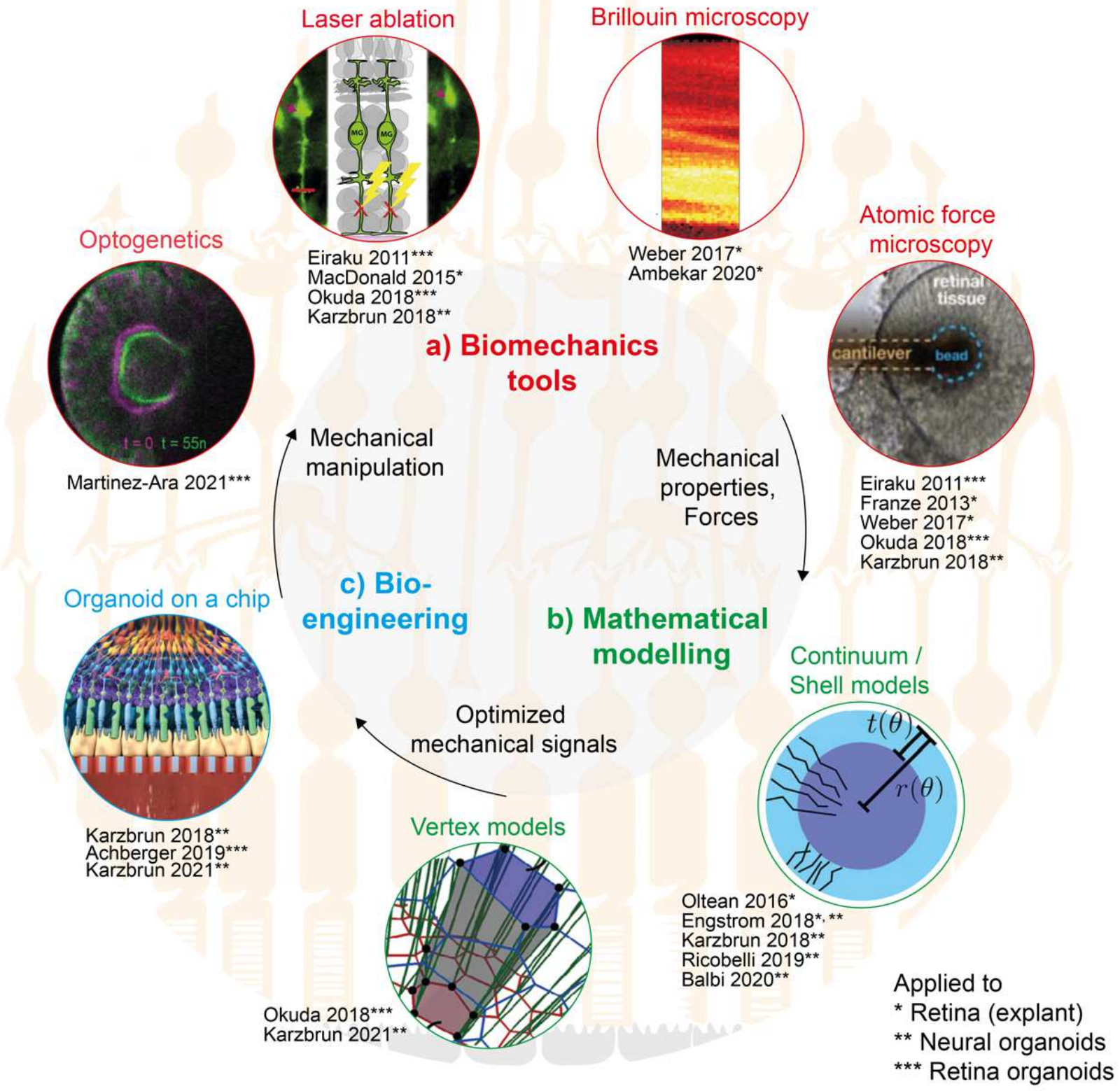

*Figure 2:* ***Towards a biophysical understanding of ROs*. *a) Biomechanics tools*** *allow the mechanical manipulation and quantification of mechanical properties and forces. Examples are stiffness measurements via Brillouin microscopy, tension characterization via laser ablation, activation of cellular forces via optogenetics as well as elastic modulus measurements via the AFM. Those measurements serve as inputs into* ***mathematical models b)*** *at different coarse graining levels. Shell and continuum models operate with a minimal set of parameters by assuming bulk material properties. The behavior of individual cells is captured in 3D vertex models. Insights from mathematical modelling now allows the* ***bioengineering of a mechanical niche c)*** *to optimize organoid growth and to specify mechanical signals. Reproduced from Eiraku et al., Nature 472 (2011),[6] with permission from Springer Nature; MacDonald et al., J. Cell Biol. 210, 7 (2015),[129] licensed under a Creative Commons Attribution (CC BY-NC-SA) license; Weber et al., Phys. Biol. 14, 6 (2017),[123] licensed under a Creative Commons Attribution (CC BY) license; Engstrom et al., Phys. Rev. X 8, 4 (2018),[152] licensed under a Creative Commons Attribution (CC BY) license; Okuda et al., Sci Adv. 4, 11 (2018),[119] licensed under a Creative Commons Attribution (CC BY) license; Achberger et al., eLife 8:e46188 (2019),[81] licensed under a Creative Commons Attribution (CC BY) license, Martinez-Ara et al., bioRxiv 2021.04.20.440475 (2021),[133] licensed under a Creative Commons Attribution (CC BY) license.*

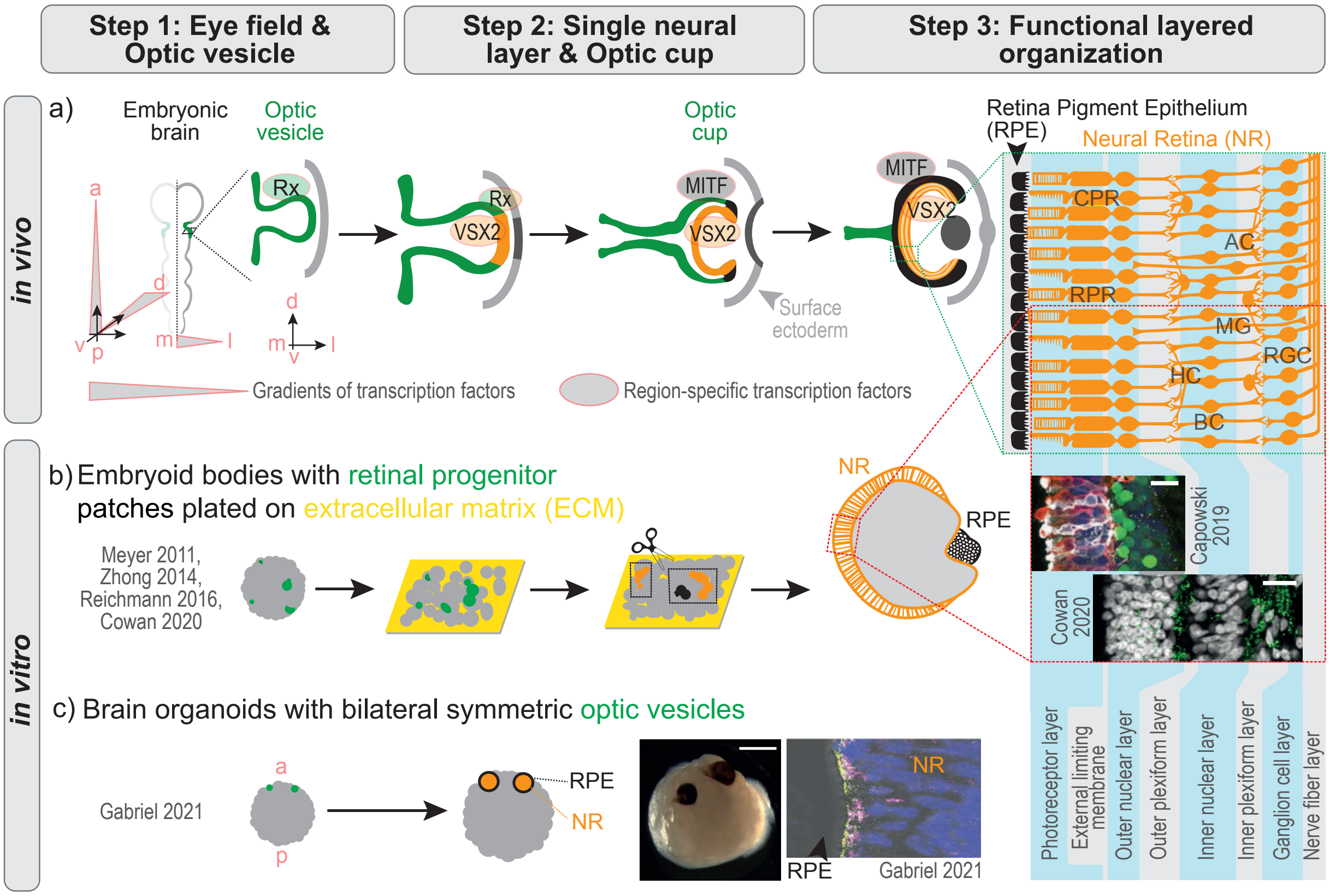
Step 1: Eye field & Optic vesicle
Step 2: Single neural layer & Optic cup
Step 3: Functional layered organization
in vivo
a)
Embryonic brain
Optic vesicle
Rx
VSX2
Optic cup
MITF
Surface ectoderm
Retina Pigment Epithelium (RPE)
Neural Retina (NR)
CPR
AC
RPR
MG
RGC
HC
BC
a
d
v
p
m
l
Gradients of transcription factors
Region-specific transcription factors
in vitro
b) Embryoid bodies with retinal progenitor patches plated on extracellular matrix (ECM)
Meyer 2011, Zhong 2014, Reichmann 2016, Cowan 2020
NR
RPE
Capowski 2019
Cowan 2020
c) Brain organoids with bilateral symmetric optic vesicles
Gabriel 2021
RPE
NR
Gabriel 2021
Photoreceptor layer
External limiting membrane
Outer nuclear layer
Outer plexiform layer
Inner nuclear layer
Inner plexiform layer
Ganglion cell layer
Nerve fiber layer

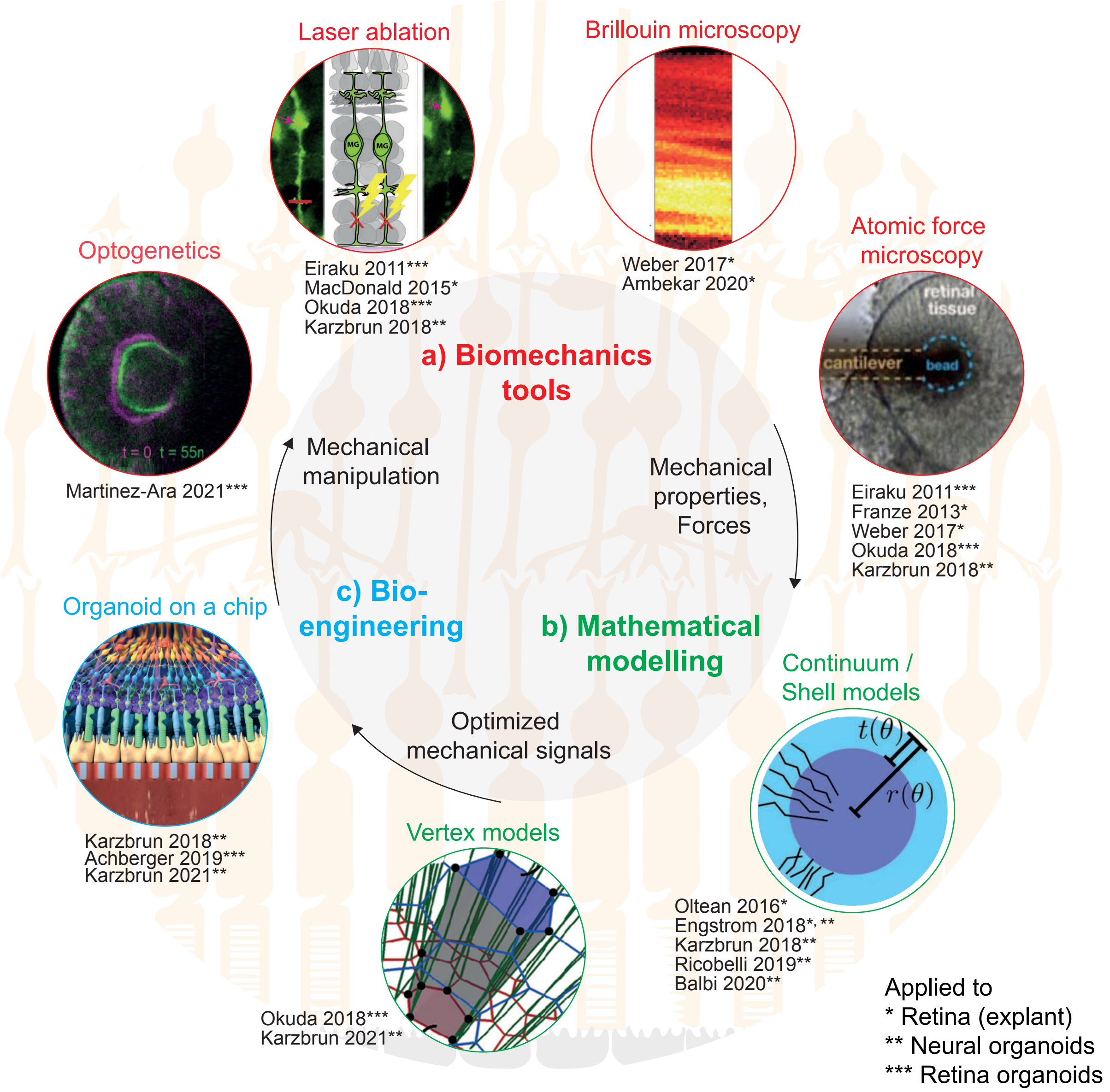